\documentclass[aip,apl,reprint,twocolumn,floatfix]{revtex4-1}

\usepackage{graphicx}
\usepackage{dcolumn}
\usepackage{bm}
\usepackage{textcomp}

\begin{document}

\title{ Spectroscopy of a Qubit Array via a Single Transmission Line }

\author{M.~Jerger}
\affiliation{Physikalisches Institut, Karlsruhe Institute of Technology and
DFG-Center for Functional Nanostructures (CFN),
D-76128 Karlsruhe, Germany}

\author{S.~Poletto}
\affiliation{Physikalisches Institut, Karlsruhe Institute of Technology and
DFG-Center for Functional Nanostructures (CFN),
D-76128 Karlsruhe, Germany}

\author{P.~Macha}
\affiliation{Institute of Photonic Technology, PO Box 100239, D-07702 Jena, Germany}

\author{U.~H\"ubner}
\affiliation{Institute of Photonic Technology, PO Box 100239, D-07702 Jena, Germany}

\author{A.~Lukashenko}
\affiliation{Physikalisches Institut, Karlsruhe Institute of Technology and
DFG-Center for Functional Nanostructures (CFN),
D-76128 Karlsruhe, Germany}

\author{E.~Il'ichev}
\affiliation{Institute of Photonic Technology, PO Box 100239, D-07702 Jena, Germany}

\author{A.~V.~Ustinov}
\email{ustinov@kit.edu}
\affiliation{Physikalisches Institut, Karlsruhe Institute of Technology and
DFG-Center for Functional Nanostructures (CFN),
D-76128 Karlsruhe, Germany}

\date{\today}

\begin{abstract}

Frequency-selective readout for superconducting qubits opens the way towards scaling qubit
circuits up without increasing the number of measurement
lines. Here we demonstrate the readout of an array of 7 flux
qubits located on the same chip. Each qubit is placed near an individual
$\lambda/4$ resonator which, in turn, is coupled to a common microwave transmission
line. We performed spectroscopy of all qubits and determined their parameters in
a single measurement run.

\end{abstract}

\pacs{03.67.Lx 85.25.Am}


\maketitle

%
%
%
%
Superconducting qubits are effective two-level quantum systems with a controllable
transition frequency between their eigenstates. They are among the most promising
candidates for registers of future quantum computers, because of their potential
to be manufactured lithographically in a controlled manner.
This gives the designer the freedom to construct
custom quantum circuits with well-defined parameters and consisting of a large
number of devices.
In practice, one of the problems that limits the scalability of qubit circuits is the readout apparatus that measures the qubit states at the end of a computation.
Traditionally, the quantum state of superconducting flux\cite{Wal-Sci-00}
or phase\cite{Martinis-PRL-02} qubits is read out by measuring the switching current of a SQUID coupled to the qubit. This readout procedure requires dedicated wiring and additional external circuitry
for every qubit. An alternative to this bulky readout is a dispersive readout realized by coupling the qubit to a superconducting resonator\cite{Ilichev-PRL-03,Wallraff-Nat-04}.
A multiplexed readout of two\cite{Majer-Nat-07} and three\cite{DiCarlo-Nat-10}
qubits through a single resonant cavity has already been demonstrated with Transmon qubits. However, the dispersive scheme cannot be scaled to a large number of devices, because one can not easily
distinguish the signals generated by different qubits.
In contrast to that, frequency-division multiplexing readout appears to be very promising.
This approach has already been demonstrated for kinetic
inductance detectors\cite{Day2003} with up to 42 devices\cite{Monfardini2010}, and it is easily extendable to measure hundreds of detectors through a single readout line.

%
%
%
%
\begin{figure}[b]
    \includegraphics{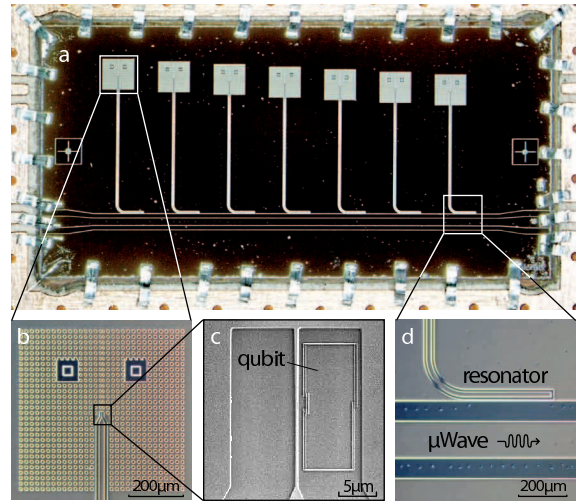}
   \caption{
   (color online) Micrographs of the sample.
	(a) Overview showing the seven quarter-wave resonators coupled to a common transmission line.
	(b) Shorted end of one of the resonators. The small holes in Nb ground plane play the role of magnetic flux traps.
	(c) One of the flux qubits.
	(d) An elbow-shaped coupling capacitor between a resonator and the feed line.
    }

    \label{fig:sample}
\end{figure}

In this Letter, we present a scalable implementation of frequency selective readout
of an array of flux qubits that uses a dedicated resonator coupled to each qubit.

The basic idea of the measurement is as follows. Due to the coupling to its qubit, each resonator acquires a dispersive shift\cite{Wallraff-Nat-04},
\begin{equation}
  \Delta\omega_r = - \frac{\tilde{g}^2}{\omega_q - \omega_r} \sigma_z,
  \label{eq:shift}
\end{equation}
depending on the state of the qubit. Here $\tilde{g}$ is the effective coupling between the resonator and qubit, 
$\hbar\omega_q$ is the transition energy between the qubit states,
$\omega_r/2\pi$ is the resonance frequency of the uncoupled resonator and $\sigma_z$
is $\pm 1$ depending on the state of the qubit. Thus the state of the qubit can be determined by measuring $\Delta\omega_r$. All resonators are coupled to a common transmission line through which their resonance frequencies can be measured.

In this experiment, we used coplanar $\lambda/4$-resonators with resonance frequencies ranging from 9.3 GHz to 10.3 GHz. These resonators were capacitively coupled to a common transmission line on one end and shorted to ground at the other end, see Fig.~\ref{fig:sample}. Outside the bands of the resonators, electromagnetic waves propagate freely along the feed line. Close to resonance, every resonator acts as a notch filter, suppressing transmission through the line.
The width of the resonance dip is determined by the internal losses of the resonator, described by the unloaded quality factor $Q_0$.
This feature offers a great advantage over using transmission-type resonators, where the line width is determined by both the losses through the  coupling capacitors and the internal loss.
With an absorption-type resonator, small dispersive shifts of the resonance frequency, induced by the qubit, can be easily detected even in the case of strong resonator coupling to the feed line.
In our samples, typical external quality factors $Q_\mathrm{ext}$ (determined by the coupling capacitors) are about $1500$, while unloaded quality factors $Q_0$ are about $40000$, resulting in half-power line widths of about 400\,kHz at 10\,GHz.

The resonators and the readout line were fabricated by e-beam lithography and CF$_{4}$ reactive-ion etching of a 200 nm thick Nb-film deposited on an undoped silicon substrate.
The aluminum flux qubits were deposited in the gap of each resonator near its shorted end by using conventional two-angle shadow evaporation technique (Fig. \ref{fig:sample}(b,c)).
They are galvanically decoupled from the resonator, making the qubit fabrication independent of the resonator and providing flexibility in design and fabrication of more complicated circuits. The qubit loop ($7 \times 16\,\mathrm{\mu{}m}^2$) is inductively coupled to the resonator, and is interrupted by three Josephson junctions. Two of them have a nominal size of $700 \times 200\,\textrm{nm}^2$, while the third junction, called $\alpha$-junction, is about 30\% smaller.

%
%
%
%
\begin{figure}[t]
    \includegraphics{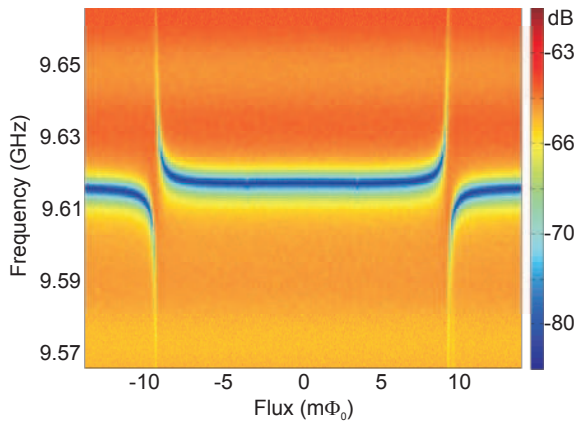}
    \caption{
	(color online) Anticrossings of qubit \#3 and its resonator.
    }
    \label{fig:anticrossing}
\end{figure}

The sample was placed in a dilution refrigerator with a base temperature of 30\,mK. The amplitude and phase of transmission through the feed line were measured using a vector network analyzer. In order to obtain a good signal to noise ratio, an amplification chain consisting of a cryogenic amplifier and two room-temperature amplifiers with a total gain of 80\,dB was used. Back-action of the cold amplifier on the system was reduced by placing isolators at the millikelvin stage of the refrigerator.

The transition frequency between the ground and first excited states of a flux qubit is given by\cite{Mooij-Sci-99}
\begin{equation}
	\omega_q = \sqrt{\Delta^2+\epsilon^2}.
\label{eq:freq}
\end{equation}
Here $\Delta/2\pi$ is the minimum transition frequency between the ground and first excited states,
at the symmetry point of the flux qubit.
The energy bias $\epsilon = 2 I_p(\Phi-\Phi_0/2)/\hbar$, expressed here in frequency units, is caused by the external magnetic flux $\Phi$ applied to the qubit loop. Thus, the transition frequency can be controlled by changing the external magnetic field applied to the qubit.

For $n$ photons in the resonator, the eigenenergies of the coupled qubit-resonator system can be expressed in the analytical form\cite{Ble04}:
\begin{equation}
  \frac{\Delta E_\pm}{\hbar} =\left(n+\frac{1}{2}\right)\omega_r+\frac{\omega_q}{2}\pm
  \sqrt{\left(\frac{\delta}{2}\right)^2+{\tilde{g}^2(n+1)}} \;,
  \label{Eq:E}
\end{equation}
where $\delta = \omega_{q}-\omega_{r}$ is the qubit-resonator detuning and $\tilde{g} = g\cdot\Delta/\sqrt{\epsilon^2+\Delta^2}$ with the bare coupling $g$. From
Eq.~(\ref{Eq:E}) it follows that the coupling leads to a change in the system's eigenenergy, revealing an anticrossing in the vicinity of $\delta = 0$. These anticrossings can be resolved by measuring transmission through the feed line.
The experimentally measured amplitude of the transmitted signal is shown in Fig.~\ref{fig:anticrossing}, where we observe two anticrossing points at $\omega_q=\omega_r$.
Note that anticrossings were already reported for the signal transmitted through coplanar $\lambda/2$-resonators\cite{Abdumalikov2008,Oelsner2010,Niemczyk2010} coupled to flux qubits.
%
%
%
%
\begin{figure}[t]
    \includegraphics{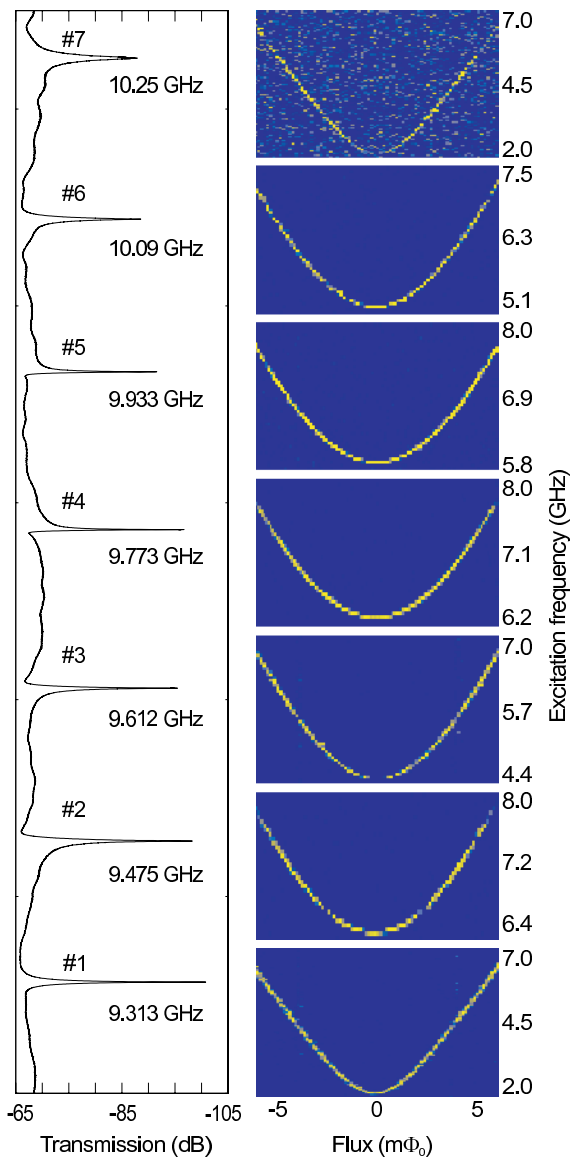}
    \caption{
	(left) Transmission spectrum of the sample, showing absorption dips of the seven resonators at zero flux bias.
	(right, color online) Spectroscopy of the corresponding seven flux qubits close to their respective symmetry points. Flux scales are shown relative to the symmetry points of the qubits.
    }
    \label{fig:spectra}
\end{figure}

\begin{table}[t]
  \newcolumntype{q}[1]{>{\raggedleft\hspace{0pt}}p{#1}}
  \begin{tabular}{cq{1.6cm}q{1.6cm}q{1.6cm}q{1.6cm}}
  \hline
  \multicolumn{1}{c}{qubit no.} & \multicolumn{1}{r}{$\omega_r/2\pi$} & \multicolumn{1}{r}{$\Delta/2\pi$} & \multicolumn{1}{r}{$g/2\pi$} & \multicolumn{1}{r}{$I_{\mathrm{p}}$} \tabularnewline
  & \multicolumn{1}{r}{[GHz]} & \multicolumn{1}{r}{[GHz]} & \multicolumn{1}{r}{[MHz]} & \multicolumn{1}{r}{[nA]} \tabularnewline
  \hline
  1 & 9.313 & 1.97 &  51 & 159 \tabularnewline
  2 & 9.475 & 6.35 &  73 & 120 \tabularnewline
  3 & 9.612 & 4.40 &  80 & 134 \tabularnewline
  4 & 9.773 & 6.17 &  85 & 126 \tabularnewline
  5 & 9.933 & 5.80 &  89 & 129 \tabularnewline
  6 & 10.09 & 5.10 &  89 & 132 \tabularnewline
  7 & 10.25 & 2.00 &  51 & 163 \tabularnewline
  \hline
 \end{tabular}
 \caption{
    Qubit gaps $\Delta$, bare coupling strengths $g$ and persistent currents $I_p$
    extracted from measurements.}
 \label{tab:qubits}
\end{table}

In order to reconstruct the qubit parameters, we carry out spectroscopic measurements in the following way. Two microwave signals are applied to the feed line: One excites the qubit and thereby changes the dispersive shift of the resonator frequency due to the qubit state. Another one weakly probes the resonator frequency. Because we designed relatively low external quality factors
$Q_\mathrm{ext}$ of the resonators, corresponding to a large coupling
capacitor between the transmission line and resonators, we can apply the spectroscopy
tone directly to the line that feeds the resonators and do not need any additional line
to manipulate the qubits.

In devices \#1--\#6, the observed frequency shift is several (unloaded) line widths, yielding a
difference in transmission between the two qubit states of well above 10\,dB.
Device \#7 has a lower $Q_0$ of 5000, yielding a difference of less than 3\,dB.

Using the technique described above, we measured the spectra of all seven qubits. The readout signal was kept sufficiently small to achieve an average photon number in the resonators below unity. This condition was verified by ac Zeeman shift measurements (not shown).

The spectra, shown in Fig.~\ref{fig:spectra}, contain important fabrication parameters of the qubits.  We combine spectroscopy data with the dispersive shift at a chosen
bias point and make use of Eq.~(\ref{eq:shift}) to extract the bare coupling $g$ between each resonator and qubit.
The persistent current $I_p$ circulating in the qubit loop is obtained from a fit of the transition
frequency vs. flux bias curve using Eq.~(\ref{eq:freq}). The measured qubit parameters are summarized in Table~\ref{tab:qubits}.
Note that all measurements were done in a single run.

Table \ref{tab:qubits} shows that the reproducibility of the qubit parameters is rather good. For the qubits in the middle of the chip (\#2--\#6) the spread of the persistent current values is below 5\%. The spread of the qubit gap is higher (above 20\% - due to its exponential dependence on the critical current of the $\alpha$-junction). Remarkable is also the relatively strong coupling $g$ (see Table ~\ref{tab:qubits}), which is due to purely inductive interaction. The deviations of the parameters of qubits \#1 and \#7 (especially coupling which differs by almost 50\% compared to the others) are currently under investigation.

%
%
%
In conclusion, we have demonstrated a frequency multiplexing readout for superconducting qubits.
Using this approach, we performed spectroscopy on seven flux qubits
by using just one on-chip transmission line. The reported readout scheme is not limited to flux qubits and can easily be implemented for other types of superconducting qubits.

%
%
This work was supported by the EU project SOLID, the Deutsche Forschungsgemeinschaft (DFG), and the State of Baden-W\"urttemberg through the DFG-Center for Functional Nanostructures (CFN) within subproject B3.4.

\end{document}